\documentclass{aa}  
\usepackage{graphicx}
\usepackage{txfonts}
\usepackage{color}
\usepackage{xspace}
\usepackage{siunitx}
\usepackage{amsmath}
\usepackage{amssymb}
\usepackage{xcolor}
\usepackage{dashbox}
\usepackage{framed}
\usepackage{lipsum}
\usepackage{placeins}
\usepackage{stfloats}

\usepackage{hyperref}

\hypersetup{
        colorlinks=true,                                                
        breaklinks=true,
        linkcolor=blue,                                                     
        citecolor=blue,                                                         
        filecolor=blue,                                                     
        urlcolor=blue,
        unicode=false,                                                      
        pdftoolbar=true,                                                    
        pdfmenubar=true,                                              
        pdffitwindow=false,                                                 
        pdfstartview={Fit},                                                 
        pdftitle={A census of the near-by Pisces-Eridanus stellar stream},  
        pdfauthor={Siegfried R\"oser},                        
        pdfsubject={},                                          
        pdfcreator={Siegfried R\"oser},                       
        pdfkeywords={}, 
        pdfnewwindow=true,                                                  
        pdfdisplaydoctitle=true                                     
}

\makeatletter
\renewcommand*\aa@pageof{, page \thepage{} of \pageref*{LastPage}}
\makeatother

\definecolor{siggi}{rgb}{0.86, 0.08, 0.24}

\definecolor{elena}{rgb}{0.01, 0.75, 0.24}

\newcommand{\gaia}{\textit{Gaia}\xspace}
\let\plotone\includegraphics

\begin{document}

   \title{A census of the near-by Pisces-Eridanus stellar stream \thanks{The full Tables 1 and 2 are only available in electronic form at the CDS via anonymous ftp to cdsarc.u-strasbg.fr (130.79.128.5) or via http://cdsweb.u-strasbg.fr/cgi-bin/qcat?J/A+A/}}
   
\subtitle{Commonalities with and disparities from the Pleiades}
       \author{Siegfried R\"oser
         \inst{}
          \and
          Elena Schilbach
          \inst{}
          }
\institute
{Zentrum f\"ur Astronomie der Universit\"at
Heidelberg, Astronomisches Rechen-Institut, M\"{o}nchhofstra\ss{}e 12-14, 69120 Heidelberg, Germany\\
    \email{roeser@ari.uni-heidelberg.de, elena@ari.uni-heidelberg.de}}

   \date{Received 9 February 2020; accepted .}


\abstract
   {}
{Within a 400~pc sphere around the Sun, we search for members of the
Pisces-Eridanus (Psc-Eri) stellar stream in the Gaia Data Release 2 (DR2) data set. We
compare basic astrophysical characteristics of the stream with those of the Pleiades.}
{We used a modified convergent-point method to identify stars with 2D - velocities
consistent with
the space velocity of the Psc-Eri stream and the Pleiades, respectively.}
{We found 1387 members of the Psc-Eri stream in a $G$ magnitude range from 5.1~mag to
19.3~mag at distances between 80 and 380~pc from the Sun. The stream has a nearly
cylindrical shape with length and thickness of about 700 pc and 100 pc, respectively.
The total stellar mass contained in the stream is about 770~M$_\odot$, and the members
are gravitationally
unbound. For the stream we found an age of about 135 Myr. In many astrophysical
properties Psc-Eri is comparable to the open cluster M45 (the Pleiades): in its age, its
luminosity function (LF), its Present-day mass Function (PDMF) as well as in its total
mass.
Nonetheless, the two stellar ensembles are completely unlike in their physical
appearance. We
cautiously give two possible explanations for this disagreement: (i)
the star-formation
efficiency in their parental molecular clouds was higher for the Pleiades than for Psc-Eri
or/and (ii) the Pleiades had a higher primordial mass segregation immediately after the
expulsion of the molecular gas of the parental cloud. } 
   {}


   \maketitle
%
%
\section{Introduction}\label{intro}

Recently, \citet{2019A&A...622L..13M} reported about the revelation of a highly elongated, coeval stream of stars in the Solar neighbourhood. Their search was based on a subset of the catalogue from Gaia Data Release 2 (DR2) \citep{2018A&A...616A...1G}, which contains stars with radial velocities measured by Gaia. The Gaia Radial Velocity Spectrometer \citep{2018A&A...616A...5C} provides measurements only for stars with surface temperatures between 3550 and 6900~K \citep{2019A&A...622A.205K}. They found 256 co-moving stars with 3D-velocities to be members in the stream. At a length of some 400 pc, the width of the stream  amounts to only about 50 pc. The 3D velocity dispersion of the members is found to be only 1.3 ~km~s$^{-1}$. 
The members follow a well-defined main sequence in the Colour-Absolute Magnitude Diagram (CAMD), and the authors estimated an age of about 1 Gyr for the stream based on giant stars among their members. Indeed, this estimate was essentially based on the presence of the G8IV sub-giant \object{42 Cet A}, which is a  part of a triple stellar system.

\citet{2019A&A...622L..13M} did not give a name to the stream, but a few months later \citet{2019AJ....158...77C} made a new age determination of the stream and named it Pisces-Eridanus (PSC-Eri) stellar stream, and we will adhere to this name throughout this paper. 
They used rotation period measurements by the \emph{Transiting Exoplanet Survey Satellite} TESS \citep{2015JATIS...1a4003R} which were available for 101 stars of the \citet{2019A&A...622L..13M} sample. From comparisons of the rotation periods of 3 benchmark open clusters, NGC 6811, Praesepe and the Pleiades, they found remarkable coincidence of the Psc-Eri stars with Pleiades stars in the (Rotation Period-Effective Temperature)-diagram which was a clear hint that both stellar aggregations are coeval. Therefore they assigned the Psc-Eri stream an age of  $\approx$ 120 Myr, i.e. much younger than estimated by \citet{2019A&A...622L..13M}. 

The first aim of this paper is to perform a membership census of the stream in 5 dimensions of phase space down to the limiting magnitude of Gaia DR2. We use this census to re-determine the age of the stream, as well as to derive its luminosity- and Present-day mass functions (LF and PDMF). As a second aim we compare its properties with those of the Pleiades. 

The paper is structured as follows:
In Section \ref{detect} we describe the steps we undertook to find the members of the Psc-Eri stream. In Section \ref{prop} the properties of the stream are discussed. We present the CAMD, the LF and PDMFs, describe the structure and kinematics of the stream and compare these properties with those of the Pleiades. In Section \ref{evo} we propose an explanation why the two stellar ensembles, the Psc-Eri stream and the Pleiades, despite of their parity in many astrophysical parameters, have such a different physical appearance. A summary concludes the paper.
\section{Finding candidate members in the Pisces-Eridanus stream}\label{detect}
For our search we used an astrometrically and photometrically clean sample of 8,131,092 stars  from Gaia DR2 with parallaxes greater than 2.4 mas, where we strictly followed the recommendations by Lindegren given in document \texttt{GAIA-C3-TN-LU-LL-124-01}, which can be found on ESA's webpage\footnote{
\texttt{https://www.cosmos.esa.int/web/gaia/}\newline  \texttt{public-dpac-documents}}.  This sample (hereafter 2.4\,mas-Catalogue) is described in \citet{2019A&A...621L...2R}; it gets incomplete at the faint end of the magnitude distribution near $G$~=~18~mag and also near the parallax limit of 2.4~mas. In the following we applied the procedures to detect overdensities in coordinate and velocity space developed in \citet{2019A&A...627A...4R}. The method is briefly repeated below.

As a starting point we determined the 6-D phase space coordinates of the approximate centre of the stream on the basis of positions, proper motions, parallaxes, and radial velocities from Gaia DR2. We used the information on membership from \citet{2019A&A...622L..13M}. The coordinates are given in the Galactic Cartesian coordinate system with origin in the  barycentre of the Solar system. The $X$-axis points to the Galactic centre, the $Y$-axis in the direction of Galactic rotation, and $Z$-axis to the Galactic north pole. The corresponding velocity coordinates are $U, V, W$. We used  232 stars from \citet{2019A&A...622L..13M} present in our 2.4\,mas-Catalogue and found mean values
\begin{equation}
\begin{array}{lcl}
\vec{R_c}  =  (X_c,Y_c,Z_c) & = & (-40,39 , +16.20, -100.10)\,{\rm pc}, \\
\vec{V_c}  =  (U_c,V_c,W_c) & = & ( -8.84, -4.06, -18.33)\,{\rm km\,s^{-1}}.\label{COPO}
\end{array}
\end{equation}
The standard deviations of the  velocity coordinates are (2.2, 1.3, 1.7)~km~s$^{-1}$.
\subsection{Constraining the range in space and velocity}\label{CVspace}
In general, a stellar stream reveals itself as an over-density in position and velocity space. \citet{2019A&A...622L..13M}
restricted themselves to stars having radial velocities measured in Gaia DR2. However, accurate radial velocities are lacking for the vast majority of stars in Gaia DR2, and they are only available for stars with effective temperatures between 3550 bis 6900 K. Therefore, we had to rely on criteria that are solely based on their tangential velocities. This implies that we may detect stars that are highly probable to be co-moving, although when the radial velocities are measured they will need final confirmation. We followed the formalism of the convergent-point (CP) method as described in \citet{2009A&A...497..209V}, for instance, and transformed the Cartesian velocity vector of the cluster motion \vec{V_c} from Eq.\ref{COPO} to give predicted velocities $V_{\parallel pred}$ and $V_{\bot pred}$ parallel and perpendicular to the direction to the CP for each star depending on its position on the sky.
This approach is appropriate and very successful to find near-by open star clusters with fixed \vec{V_c} and low velocity dispersion. Yet, the Pisces-Eridanus stream from \citet{2019A&A...622L..13M} extends over more than 400\,pc in space, and the assumption of a fixed \vec{V_c} may not be justified. In fact, we find high correlations between the spatial coordinates $X$ and $Y$, and the velocity components $U$ and $V$ of the 232 stars from \citet{2019A&A...622L..13M}. The correlation coefficients are 0.90 for $U$ vs. $Y$ and 0.66 for $V$ vs. $X$.
 \begin{equation}
\begin{array}{ll}
  U = 0.031 \times Y -9.35,\,\,\,   V = 0.015 \times X -3.46.\label{corr}
\end{array}
\end{equation}
There is no considerable correlation of the velocity component $W$ with spatial coordinates. The correlations in Eq.\ref{corr} are an intrinsic property of the \citet{2019A&A...622L..13M} sample of stars in the Psc-Eri stream. So, we cannot simply follow the classical convergent point algorithm which uses a constant space velocity such as \vec{V_c} from Eq.\ref{COPO} in order to find new candidate members of this stream. Instead, we calculate the predicted velocities $V_{\parallel pred}$ and $V_{\bot pred}$ for a star with cartesian coordinates $X,Y,Z$ by using $U,V$ from Eq.\ref{corr} and $W$ from Eq.\ref{COPO}. 

Also with this modification to the convergent point method $V_{\bot pred} \equiv$  0 still holds. Next, we similarly transformed the measured (observed) tangential velocities for each star, i.e. $\kappa\,\mu_{\alpha*}/\varpi$ and $\kappa\,\mu_{\delta}/\varpi$ into $V_{\parallel obs}$ and $V_{\bot obs}$. In this case $\varpi$ is the measured trigonometric parallax in Gaia DR2 and $\kappa=4.74047$ is the transformation factor from 1~mas~yr$^{-1}$ at 1~kpc to 1~km~s$^{-1}$. We set in the following 
$\Delta V_{\parallel}$ = $V_{\parallel obs}$ -  $V_{\parallel pred}$ and $\Delta V_{\bot}$ = $V_{\bot obs}$ -  $V_{\bot pred}$.
We also determined the covariance matrix for the velocities $V_{\parallel obs}$ and $V_{\bot obs}$ via error propagation from
the covariance matrix of the $\mu_{\alpha*}, \mu_{\delta}$, and $\varpi$.

For further analysis we did not use the full 2.4\,mas-Catalogue. We searched in a sphere of 350 pc around \vec{R_c} from Eq.\ref{COPO} and additionally constrained this volume (surrounding the Pisces-Eridanus stream) by a cut in the z-plane as $|{Z - Z_c}| \leq 100$~pc, where $Z_c$ is given in Eq. \ref{COPO}.  We also introduced a restriction in the tangential velocity plane by requiring $|\Delta V_{\parallel}|\leq $4~km~s$^{-1}$ and  $|\Delta V_{\bot}| \leq$4~km~s$^{-1}$, a total area of 64~${(\rm{km~s}^{-1})}^2$. These cuts are sufficient to find the members of the Psc-Eri stream. The restrictions reduced our selection to a subsample containing 42,733 stars which populate this search volume. 

%
\subsection{Method to find over-densities}\label{ITT} 
The method is described in detail in \citet{2019A&A...627A...4R}. We repeat here only the basic idea. The aim of the method is to separate over-densities such as clusters or moving groups (signal) from the local Galactic background (noise). For each star $i$ out of our sample of 42,733 stars in the previous section we define a neighbourhood in physical and velocity space via the following definition: a star $j$ with 5-D coordinates $(X_j,Y_j,Z_j,\Delta V_{\parallel_{j}},\Delta V_{\bot_{j}})$  is a neighbour to star $i$ with coordinates $(X_i,Y_i,Z_i,\Delta V_{\parallel_{i}},\Delta V_{\bot_{i}})$ if 
\begin{equation}
(X_i-X_j)^2 + (Y_i-Y_j)^2 +(Z_i-Z_j)^2 \leq {r_{lim}}^2,
\end{equation} 
and
\begin{equation}
{(\Delta V_{\parallel_{i}}-\Delta V_{\parallel_{j}})^2} + {(\Delta V_{\bot_{i}}-\Delta V_{\bot_{j}})^2} \leq {a^2}.
\end{equation}
The free parameters $a$ and ${r_{lim}}$ can be specified according to the goal of the study. In this paper we chose $a$ = 1.0~km~s$^{-1}$ and  r$_{lim}$ = 10~pc. When a star $i$ is surrounded by $k$ neighbours, we call this case a "$k$-neighbourhood". In our sample of 42,733 stars we found 33,509 stars belonging to a zero-neighbourhood, 3,577 to a one-neighbourhood and 1,685 to a two-neighbourhood. This corresponds to 94.9\% of our sample. Assuming that the distribution of field stars follows a  Poisson distribution, \citet{2019A&A...627A...4R} calculated the probability $p_{cont}$ of field star contamination for each $k$-neighbourhood. In our sample  of 42,733 stars, zero- and one-neighbourhoods consist overwhelmingly of field stars. Stars with two neighbours have $p_{cont}$ = 0.64, i.e. 64\% contamination, while stars with 3-neighbours have only $p_{cont}$ = 0.12.
We restricted therefore our selection of member candidates of the Psc-Eri stream to stars  with at least 3 neighbours ($k \geq$ 3), which holds for 2,162 stars. For simplicity we call this sample 3NB (NeighBours) sample.  
\begin{figure*}
\begin{minipage}[t]{0.310\textwidth}\vspace{0pt}
\includegraphics[width=\textwidth]{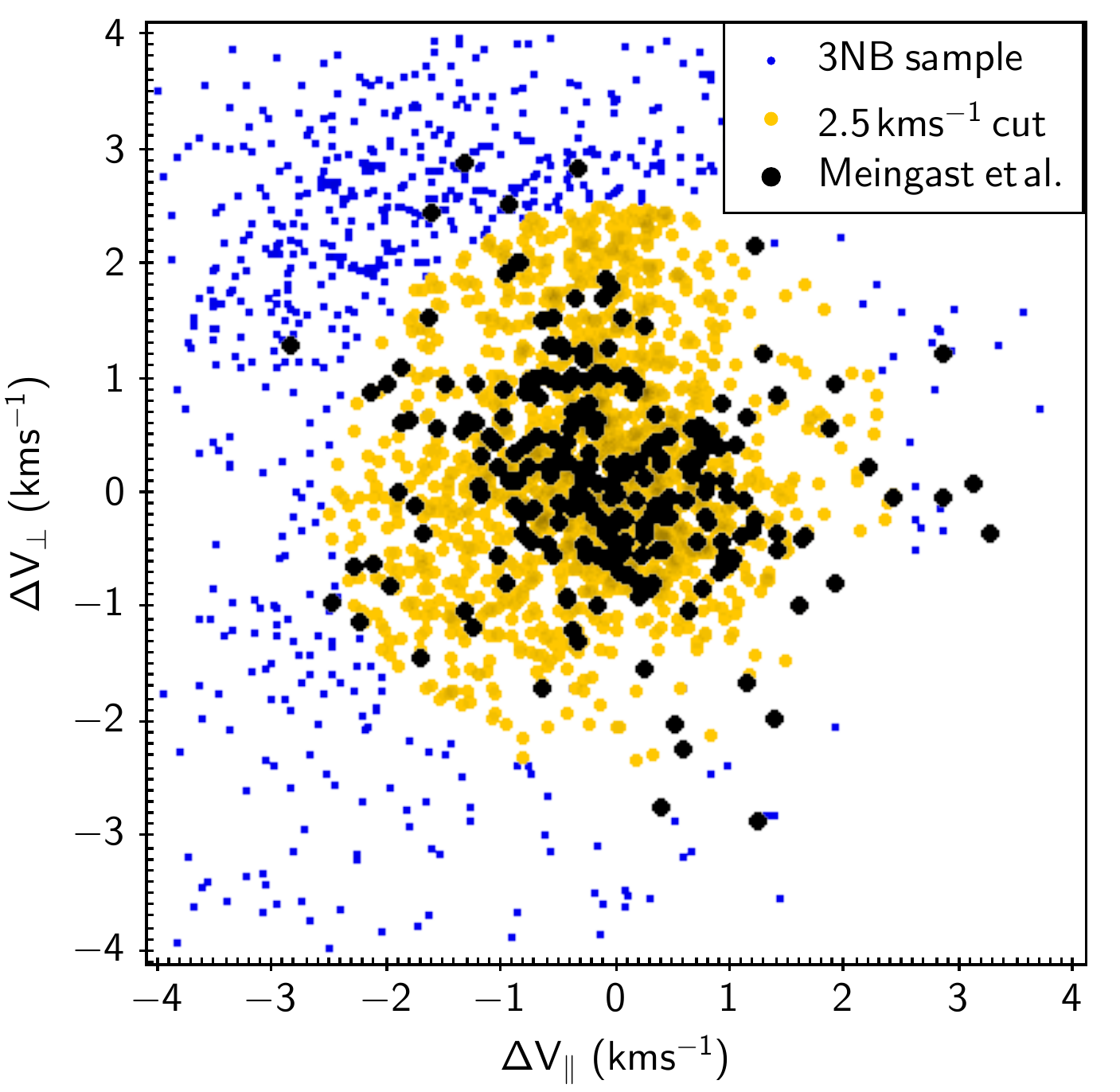}
\end{minipage}\hfill%
\begin{minipage}[t]{0.330\textwidth}\vspace{0pt}
\includegraphics[width=\textwidth]{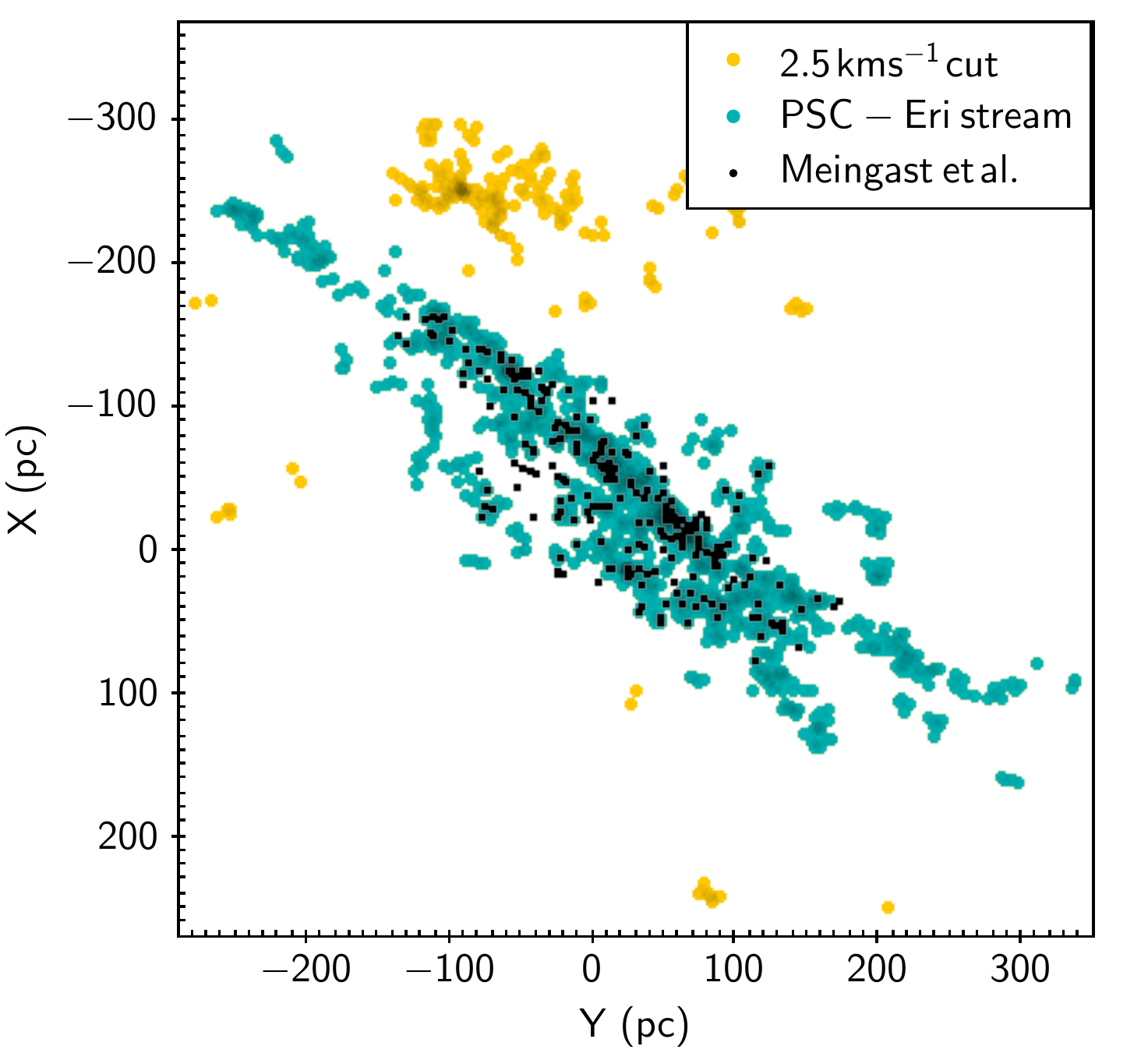}
\end{minipage}\hfill%
\begin{minipage}[t]{0.330\textwidth}\vspace{0pt}
\includegraphics[width=\textwidth]{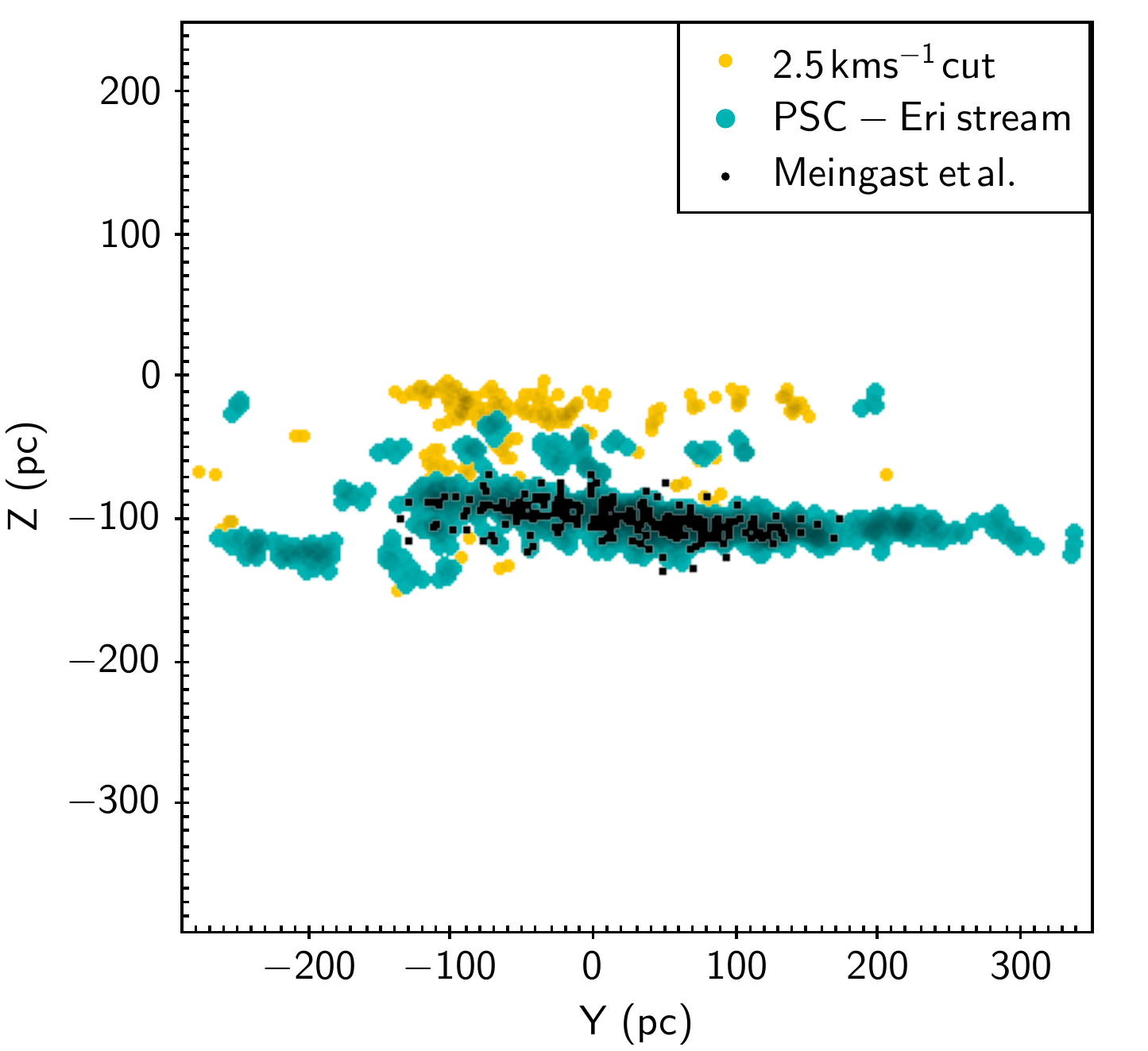}
\end{minipage}\\
      \caption{Left panel: Distribution of the Psc-Eri stars from the 3NB sample (see Sec.\ref{ITT}] shown as blue dots in the tangential velocity plane $\Delta V_{\parallel}$, $\Delta V_{\bot}$ (see text for further explanation). The yellow dots represent a subset of the 3NB sample within a radius of 2.5~km~s$^{-1}$ around the origin. The \citet{2019A&A...622L..13M} stars found in our 3NB sample are marked by black dots. Middle and right panels: Distribution of stream candidates in the $Y,X$-plane and the $Y,Z$-plane. The cyan dots show the final sample of the Psc-Eri stream. The yellows and black dots represent the same data as in the left panel.}
\label{Figure1}
\end{figure*} 
\subsection{Selection of candidates of the Psc-Eri stream}\label{candidates}
Due to the huge extend in space of the Psc-Eri stream even stars with less than 3 neighbours in a 5-D  phase space
coordinates may have been members of the stream in the past, but are now indistinguishable from field stars, 
and their  possible membership cannot be proven by kinematic arguments. Moreover, former members may have now 
velocities significantly different from the expected velocities of current stream members. That, also, makes it  difficult to distinguish them from field stars.

The left panel of Fig.~\ref{Figure1} shows the distribution of the 2162 stars (3NB sample) in the $\Delta V_{\parallel}$, 
$\Delta V_{\bot}$-plane, selected as probable stream candidates in the previous section. The loci of the velocities of the stars belonging to the Psc-Eri stream is expected to be close to the origin, and, in fact we find  a significant 
concentration of stars towards the centre of the plot. Taking velocity 
scattering of probable members from \citet{2019A&A...622L..13M} into consideration, we restricted our 
sample to stars whose observed and predicted tangential velocities differ by less than 2.5~km~s$^{-1}$. 
This cut reduces the sample to 1584 stars (2.5~km~s$^{-1}$ cut in Fig.~\ref{Figure1}, shown as yellow dots). 
 
In the next step we show the spatial distribution of these stars in the middle and right panels of 
Fig.~\ref{Figure1}  in the $Y,\, X$-plane and in  the $Y,\, Z$-plane, respectively. Although somehow fragmented,
the over-density representing the Psc-Eri stream dominates the picture, and it is clearly separated in space from 
other groups. We used 
TOPCAT\footnote{\texttt{http://www.star.bris.ac.uk/{\textasciitilde}mbt/topcat/}} \citep{2005ASPC..347...29T} to extract 
the stars belonging to the Psc-Eri stream out of the 1584 stars fitting the velocity criteria within 2.5~km~s$^{-1}$. 
One manual cut in the $X,\, Y$-plane was sufficient to give the final number of 1387 probable members (marked as cyan  dots in the middle and right panel) of the Psc-Eri stream.

By summing up $p_{cont}$ of these stars we estimated the internal contamination of the sample by 37 field stars (or 2.7\%)
where 34 of them come from the 276 stars with 3 neighbours alone. From the point of view of Poisson statistics,
the contribution to contamination by stars with 4 or more neighbours can practically be neglected in our case.

These values hold only for the contamination within the \mbox{5-D} approach. An additional contamination may come from
the fact that we did not and could not consider radial velocities for estimating membership, since radial velocity data are missing  for
the majority of stars in our sample. Nevertheless, for 239 of 1387 stars radial velocities are available in Gaia DR2 with
an accuracy better than 5~km~s$^{-1}$. Only 18 of them (7.5\%) have radial velocities inconsistent with the space velocity 
predicted for the Psc-Eri stream. By combining these two kinds of contamination, we expect a total contamination between 
about 100 to 150 stars for our final sample.

At this stage it is also appropriate to note that the sample includes those stars which fit all requirements we 
described above. Taking into account the huge size of the stream and velocity variations of its members, the sample
is by sure not complete, especially due to isolated members. Relaxing the criteria may increase the number of 
stream candidates but, the degree of field contamination will be increased, too. Therefore, we consider our sample as
a compromise, presenting, however, a basis for studies on the Psc-Eri stream. To adopt a name for this final sample of
1387 stars we simply call it Psc-Eri hereafter. The list of these stars is given in Table  \ref{tab:streamsources}.
\begin{figure*}
\begin{minipage}[t]{0.490\textwidth}\vspace{0pt}
\begin{center}
\includegraphics[width=\textwidth]{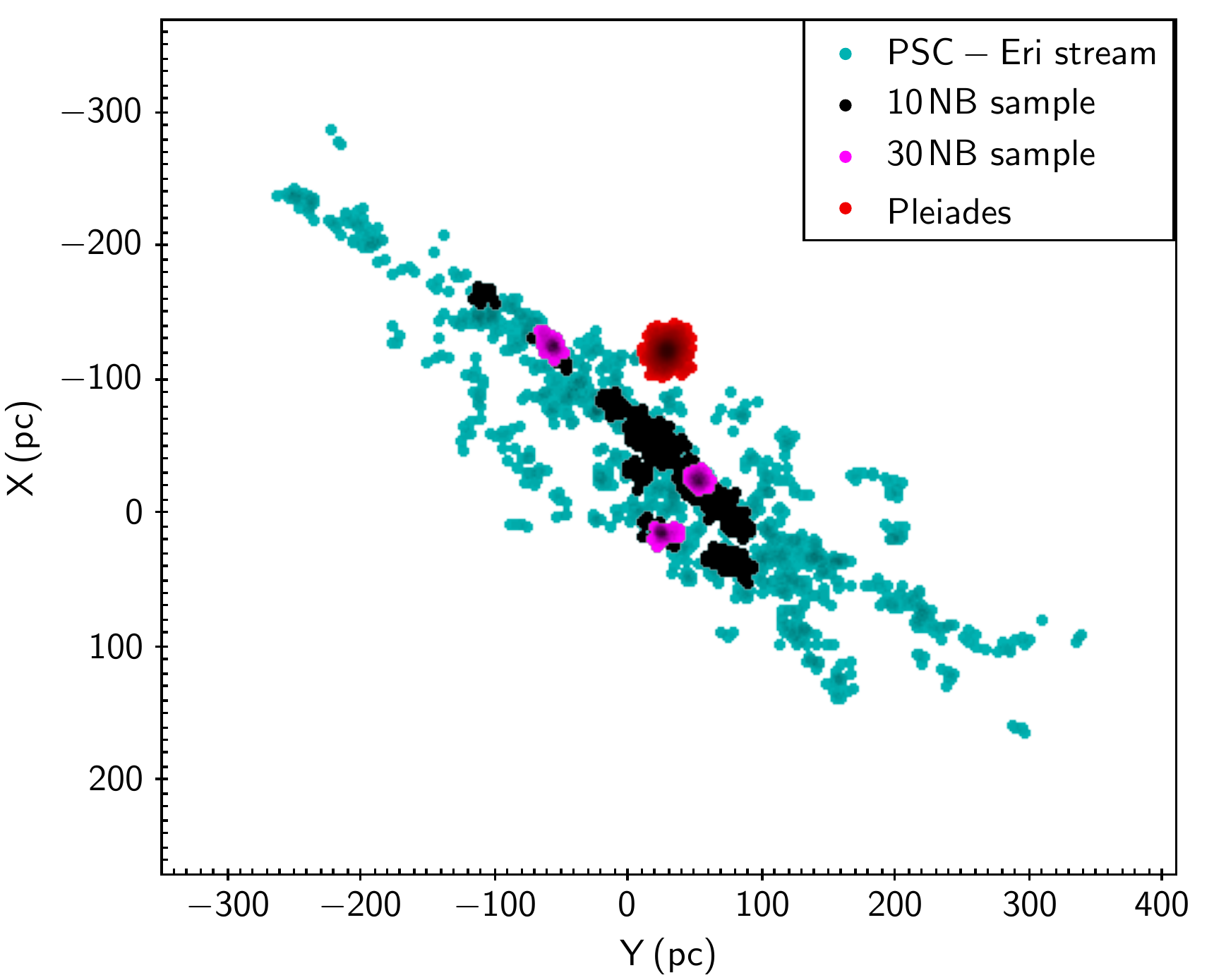}
\end{center}
\end{minipage}\hfill%
\begin{minipage}[t]{0.490\textwidth}\vspace{0pt}
\includegraphics[width=\textwidth]{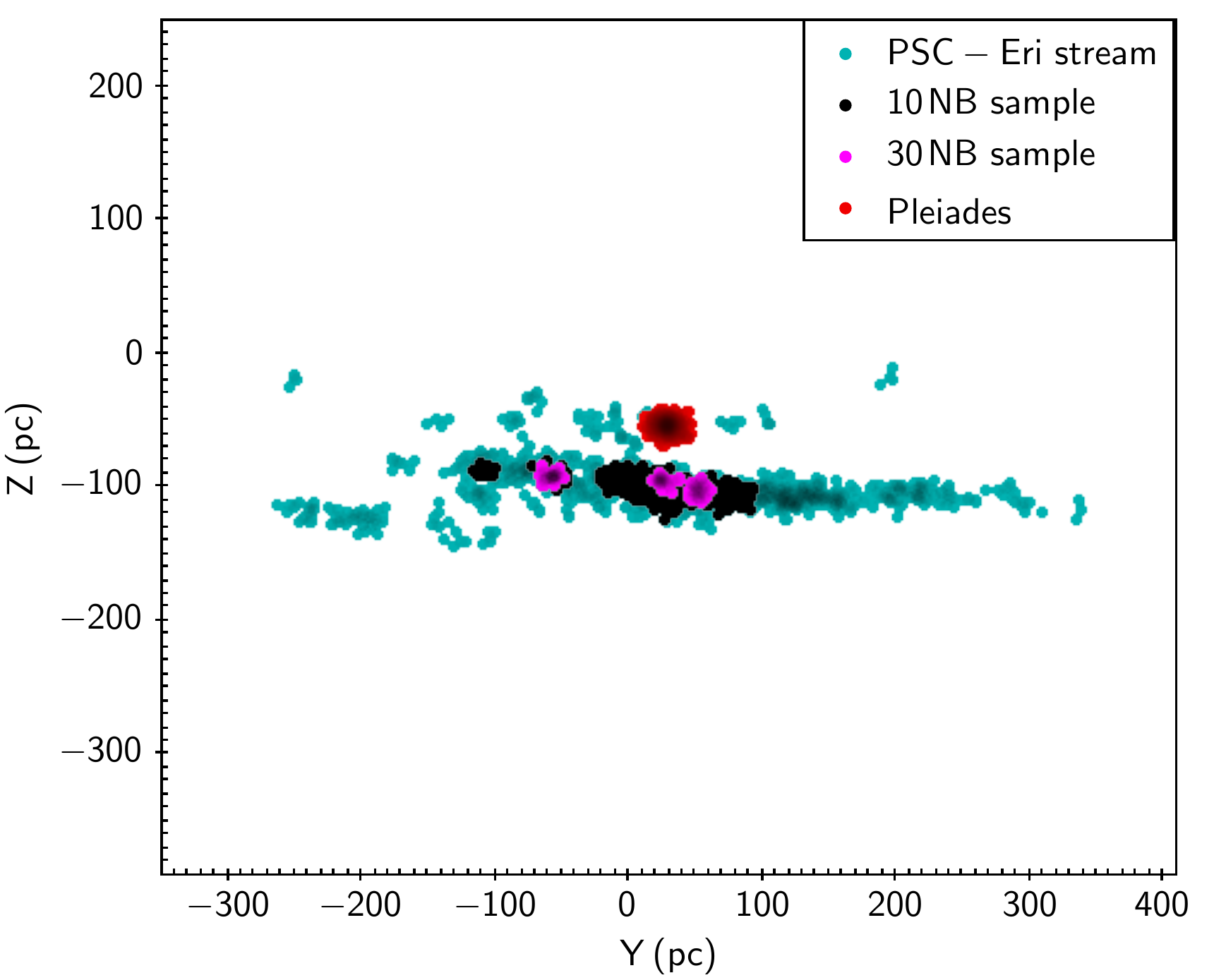}
\end{minipage}\\
      \caption{The Psc-Eri stream (cyan dots) and the Pleiades (red dots) in the $Y,X$-plane (left) and the $Y,Z$-plane (right). 
The black and pink dots indicate dense areas of the stream represented by stars with at least 10 and 30 neighbours, respectively (see Sec.\ref{Kinema}).}
\label{Figure2}
\end{figure*}
\subsection{Comparison with the \citet{2019A&A...622L..13M} membership list}\label{vsMeingast}
One would expect that the candidates from \citet{2019A&A...622L..13M} should all be included in our sample of 42,733 stars. In fact, from their 256 stars we find only 232 (90.6\%). As \citet{2019A&A...622L..13M} mentioned in their Sec.\,2, they had relaxed their selection criteria for astrometric and photometric quality. And in fact, all these 24 stars are missing in our 2.4-mas catalogue (see Sec.\ref{detect}). Note, that this does not exclude these stars as Psc-Eri members, because improved astrometric and photometric accuracy in future Gaia releases may prove their membership. Another 66 Meingast stars did not make it into our basic sample of the Psc-Eri stream, mainly because they have less than 3 neighbours, and a few are off by more than 2.5~km~s$^{-1}$ in the tangential velocity plane. Again, this does not mean that these stars are not physical members, but we cannot rule out that they are field stars. We note that we did not find \object{42 Cet A} - the object crucial for age determination in \citet{2019A&A...622L..13M} - among our member list, because it failed the strong criteria for astrometric and photometric quality.

%
\subsection{Comparison with the \citet{2019AJ....158...77C} membership list}\label{vsCurtis}
\citet{2019AJ....158...77C} added a total of 34 stars to the \citet{2019A&A...622L..13M}-sample.
They searched for candidate members, more massive than those of Meingast, via 2 different approaches. First, they selected candidates from Gaia DR2 within 5~km~s$^{-1}$ of the median value of the space velocity of the \citet{2019A&A...622L..13M} members and within 20 pc of at least one member. For this they used radial velocities found in SIMBAD \citep{2000A&AS..143....9W}. Second, they included stars within a 10\,pc volume around a Meingast star, and co-moving in accordance with the proper motion criterion $\delta(\mu)$ < 2 mas/y. This search yielded 34 stars. 
Out of these 34 high-mass member candidates, 32 are in our sample of 42,733. The 2 exceptions are HD 218242 (Gmag = 5.631) and HD  27467 (Gmag =8.999), which even did not survive the astrometric and photometric quality check. 
Out of the 32 only 15 make it into the basic sample of the Psc-Eri stream. The others are lacking because they failed our selection criteria similar to the 66 Meingast stars. On the other hand, we identified 33 new candidate members brighter than M$_G$ = 2.5\,mag, that are not in \citet{2019AJ....158...77C}. 
\section{Properties of the Psc-Eri stream}\label{prop}
Before starting the discussion on the astrophysical properties of Psc-Eri, we summarise what is known so far on the age of the stream. Using the CAMD of stream members, \citet{2019A&A...622L..13M} obtained an age of about 1 Gyr for Psc-Eri. To identify the members they  used the Gaia DR2 data for stars with radial velocities measured by the Gaia radial velocity spectrometer \citep{2018A&A...616A...5C}. These stars cover a relatively short part of the the CAMD 
($2.2 < M_G < 8.7$) and do not show a clear turn-off point that would provide a reliable age estimation. The only star which supports an age of 1 Myr is the sub-giant \object{42 Cet A}. As part of a triple system, this star turns out to be problematic for Gaia because it does not fit the astrometric and photometric quality checks. 

Recently \citet{2019AJ....158...77C} reported the age estimation of Psc-Eri based on rotation period measurements by the \emph{Transiting Exoplanet Survey Satellite} TESS \citep{2015JATIS...1a4003R}. They found that the Psc-Eri stream is coeval with the Pleiades cluster which is the benchmark for an $\approx$130-Myr old population of Galactic disk stars. Adding 34 new brighter candidates to the Meingast stars in the CAMD, \citet{2019AJ....158...77C} found that this sequence is best fitted by a 130 Myr solar-metallicity isochrone. So, both independent approaches consolidate their finding of $\approx$ 120 Myr for the age of the Psc-Eri stream. 

Given the same young age but apparently a very different visual appearance, the Pleiades and Psc-Eri provide an opportunity to compare their properties and to find out what makes them similar or distinct.

\subsection{The Pleiades sample}\label{PLE}
To make a comparison as fair as possible, we selected Pleiades members using the same method as we applied to identify the 
Psc-Eri stream in Sect.\ref{ITT} above, with only one exception. Being a compact open cluster, the Pleiades members follow
a nearly constant space motion. Therefore, we could assume a common velocity of 
\begin{align}\vec{V_c}  =  (U_c,V_c,W_c) =  ( -6.88, -28.48, -14.32)\,{\rm km\,s^{-1}}\end{align} and return to the classical CP-approach. These mean values are based on the velocity data in the membership list of \citep{2018A&A...616A..10G}. 
We chose the same selection parameters as for the Psc-Eri stream and obtained 1245 stars within a spatial cut made at 
two tidal radii (i.e. 23 pc, derived in Sec.\ref{Lum}) from the centre. The list of these stars is given in Table \ref{Pleiades}. Due to a high concentration in space and velocity, the Pleiades sample 
is practically not contaminated by field stars.  
A probable contamination due to missing information on radial velocities is of the order of 30 field stars among our 1245 Pleiades members. 

In Fig.~\ref{Figure2} we show the locations of the Pleiades and the Psc-Eri stream in the $X,Y$- (left panel) and the $Y,Z$-plane (right panel). It is amazing that, at present epoch, these two different stellar ensembles are so close together in space.
\subsection{Colour-Absolute Magnitude Diagram and age}\label{CAMDs}
\begin{figure}[hbtp]
   \centering
   \includegraphics[width=0.49\textwidth]{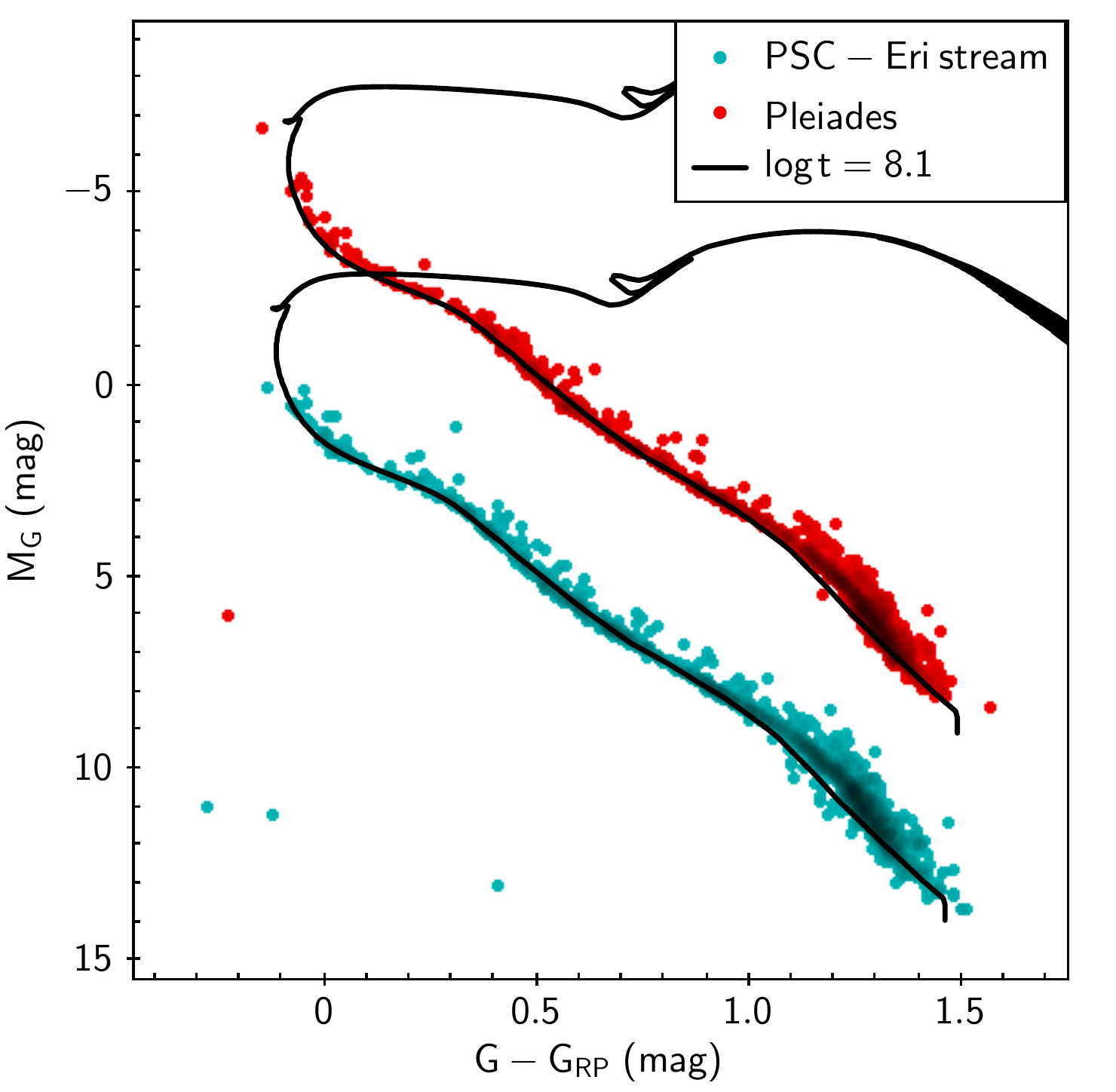}
      \caption{Colour-absolute-magnitude diagram (CAMD) $M_G$ vs. $G - G_{RP}$ of the 1387 stars of Psc-Eri (cyan dots) and of
the 1245 stars of the Pleiades (red dots). The Pleiades sequence is shifted by -5 mag along the $M_G$-axis for better visualisation. The black curves are PARSEC (version 1.2S) isochrones \citep{2014MNRAS.444.2525C} with Z = 0.02 and $\log \rm{t}$ = 8.1.}
         \label{Figure3}
   \end{figure}

In Fig.~\ref{Figure3} we show the CAMD $M_G$ versus $G - G_{RP}$ of the 1387 stars of Psc-Eri  (cyan dots) and the
1245 stars of the Pleiades (red dots). The Pleiades sequence is shifted by -5 mag along the $M_G$-axis for better visualisation. We also display a PARSEC (version 1.2S)
isochrone \citep{2014MNRAS.444.2525C} with Z = 0.02 and $\log \rm{t}$ = 8.1. For the Pleiades we took into account
a reddening of $E(B-V)$ = 0.045 mag \citep{2018A&A...616A..10G}. Since the Psc-Eri stream spreads out over a wide part
of space, the assumption of a unique reddening is not fully appropriate for the Psc-Eri sample. Nevertheless, even without
any correction for reddening the isochrone gives a reasonable fit to the observations over a wide range of $G - G_{RP}$.
An exception is a part concerning the low-mass stars redder than about $G - G_{RP}$ = 1.1 mag. However, this discrepancy
is valid for both, Psc-Eri and the Pleiades, and this is probably a calibration problem of the observations and/or isochrones.
In the Pleiades we found one white dwarf (WD) within two tidal radii from the centre, the well known \object{EGGR 25}. 
Within the Psc-Eri stream we could identify 3 WDs as probable candidate members, 
\object{KUV 03520+0500} and \object{SDSS J221945.40+032433.7}, known as WDs in SIMBAD, as well as \object{Gaia DR2 2992020285939176704} 
which is mentioned as a WD candidate in \citet{2019MNRAS.482.4570G}.

While the Pleiades are a self-gravitationally bound open cluster and the Psc-Eri stream is an unbound moving group, 
the similarity over the whole range of magnitudes of the CAMDs of the two stellar groups is striking. The CAMDs of the Pleiades
and Psc-Eri are consistent with an age between $\log \rm{t}$ = 8.0 (100 Myr) and $\log \rm{t}$ = 8.2 (150 Myr). 
Unfortunately, Gaia DR2 does not provide reliable data for stars brighter than $G \approx 6$ mag which could help to
narrow the age range from Gaia observations alone. In any case, our results  support the finding of a young age for the
Psc-Eri based on the rotation ages from the TESS satellite \citep{2019AJ....158...77C}.
\begin{figure*}
\begin{minipage}[t]{0.480\textwidth}\vspace{0pt}
\begin{center}
\includegraphics[width=\textwidth]{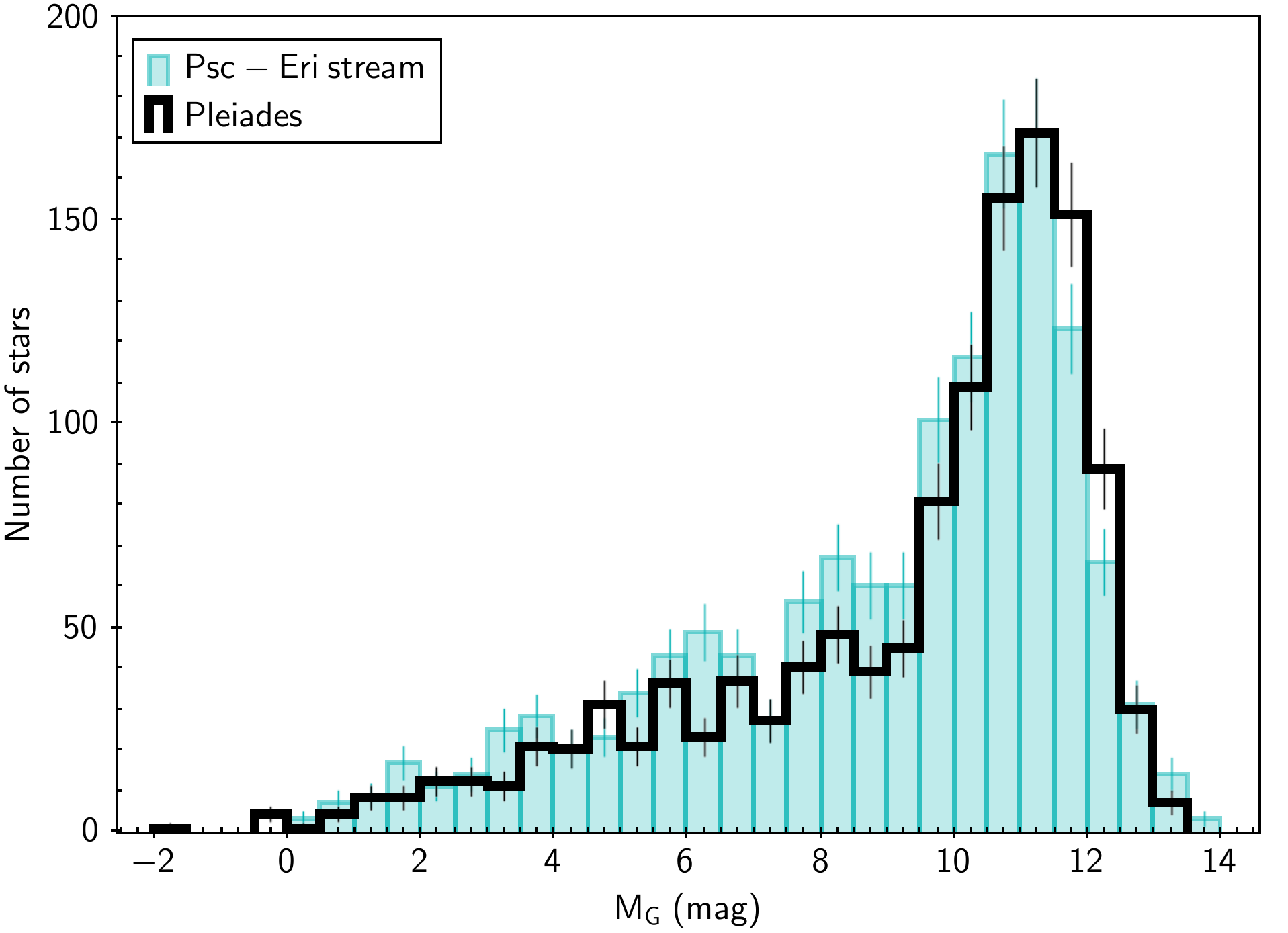}
\end{center}
\end{minipage}\hfill%
\begin{minipage}[t]{0.510\textwidth}\vspace{0pt}
\includegraphics[width=\textwidth]{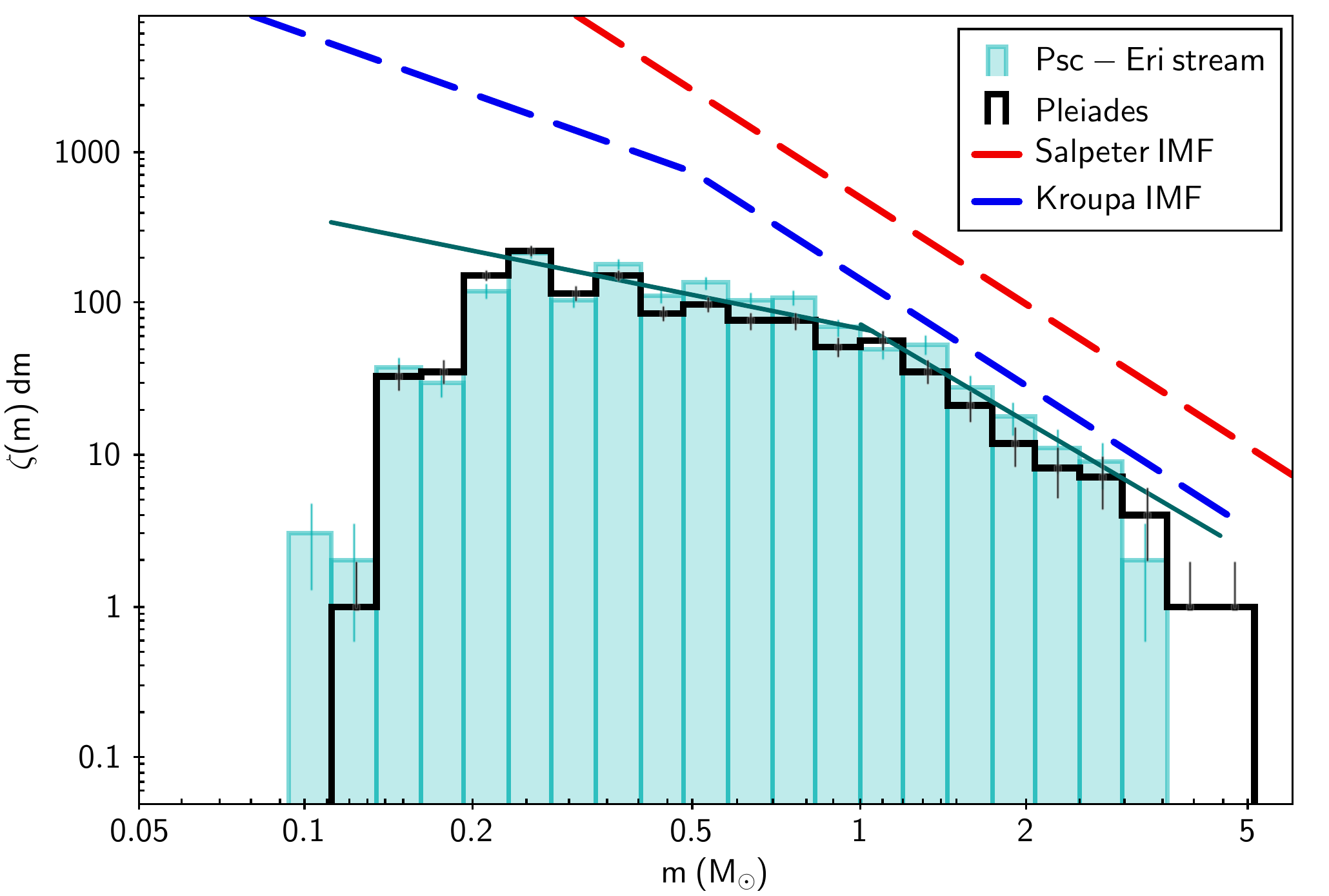}
\end{minipage}\\
      \caption{Left: Luminosity function (LF) of the Psc-Eri stream (cyan) and of the Pleiades (black steps). Right: Present-day mass function (PDMF) of the same groups, Psc-Eri stream (semi-filled cyan bars), Pleiades (black step function). The black line shows a power-law MF with a slope of -2.15 fitted to the Psc-Eri LF for a mass range between 1 and 2.5~M$_\odot$, and a slope of -0.73 in the mass range 0.3 to 1.0~M$_\odot$. The dashed red line shows an IMF with the Salpeter slope of -2.35, whereas the dashed blue line is the MF from \citet{2001MNRAS.322..231K}.}
\label{lf}
\end{figure*}
\subsection{Luminosity- and mass-functions}\label{Lum}
The luminosity functions (LF) of the
Pleiades and Psc-Eri are shown in Fig.\ref{lf} (left panel). As mentioned before, Gaia DR2 is lacking good photometric data for the apparently brightest stars. Therefore both luminosity functions may be incomplete at the bright end.
For the Pleiades one can check the completeness by comparing with well-established Pleiades membership lists based on Hipparcos, e.g. with the one used by \citet{2009A&A...497..209V}. We found that our Pleiades sample lacked 10 Hipparcos stars absolutely brighter than $M_V$ = 2.0\,mag. Similarly, in Psc-Eri comparably bright stars may also be missing. 

Given the slightly larger total number of the Psc-Eri members, we observe a little higher level in the LF for stars brighter than about $M_G = 11\,$mag. Beyond that magnitude the Pleiades are apparently higher populated. 
This is a simple consequence of the effective limiting magnitude $G \approx 18$~mag of our  2.4\,mas-Catalogue: the Psc-Eri stream extends up to 380 pc from the Sun, whereas the bulk of the Pleiades stars is located between 130 and 140 pc. 
The corresponding limiting absolute magnitudes are $M_G \approx$ 12.4\,mag for the Pleiades, $M_G \approx$ 13.5mag and\, 10.1 mag for the Psc-Eri stream at 80\,pc and  380\,pc, respectively.
Nevertheless, the luminosity functions of both stellar samples are essentially identical over a large range of absolute magnitudes.

While the luminosity function of a stellar population is based on astrometric and photometric observations alone, one needs a theoretical mass-luminosity relation to obtain masses. For Psc-Eri, as well as for the Pleiades, we used the PARSEC (version 1.2S) isochrone \citep{2014MNRAS.444.2525C} with Z = 0.02 and $\log \rm{t}$ = 8.1 shown in Fig.~\ref{Figure3}. We derived system masses of the stars via a mass-luminosity relation in all three Gaia photometric bands, neglecting binary issues. Again, the limiting magnitude of $G \approx 18$~mag yields lower mass limits of about 0.16 M$_\odot$ for the Pleiades, as well as 0.11 M$_\odot$ at 80\,pc and 0.4 M$_\odot$ at 380\,pc for Psc-Eri.

To derive the mass of the Pleiades we determined the Jacobi (or tidal) radius $r_J$ of the cluster as described in \citet{2019A&A...627A...4R}. Adding up the individual masses of the members from the centre outwards, and considering about 30~M$_\odot$ from the missing 10 absolutely bright stars we found $r_J$~=~11.5~pc. Within $r_J$ 1038 stars give a tidal mass $M_J$ = 588~M$_\odot$ for the Pleiades. Following the notation in \citet{2018ApJ...863..171S} the mass enclosed within  $r_J$ is the bound mass. Younger clusters are surrounded by an envelope of unbound stars \citep{1987ApJ...323...54E}, most of them located beyond one Jacobi radius but still within 2\,$r_J$. According to numerical simulations of the evolution of open clusters \citep{2018ApJ...863..171S}, such envelopes persist for many mega-years. We found such an envelope in the Pleiades containing more than 200 stars with a total mass of about 100~M$_\odot$. So, we call the "total" mass of the Pleiades the mass contained within two Jacobi radii, i.e. 690~M$_\odot$ in 1245 stars.

Adding up all the individual masses of the stars in the Psc-Eri stream we obtained 771~M$_\odot$. Needless to mention that the Psc-Eri stream is unbound, since the tidal radius for 771~M$_\odot$ would be 12.6\,pc. Based on the individual (system) masses we show the mass functions $\zeta(m)$ of Psc-Eri and the Pleiades in Fig.\ref{lf} (right panel), where $\zeta(m)\,dm$  gives the total number of stars in the mass range between $m$ and $m+dm$.

In a comparison of both mass functions we disregard the part for masses larger than about 2.5~M$_\odot$ where we already noted incompleteness in both samples. At the low-mass end of the Psc-Eri mass function we find a few stars in the first two bins nicely coinciding with the expected limit of 0.11~M$_\odot$ for the members closest to the Sun. Also the observed PDMF of the Pleiades strictly adheres to its expected low-mass limit of 0.16~M$_\odot$. The mass functions of both aggregations can be well described by two power laws, one between 0.3 and 1.0~M$_\odot$ and another between 1.0 and 2.5~M$_\odot$. In the latter mass-range the slope of the mass function of the Pleiades is -2.41, that of the Psc-Eri stream slightly shallower with an exponent of -2.15. Both slopes are close to the Salpeter slope \citep{1955ApJ...121..161S} of -2.35. In the mass range between 0.3 and 1.0~M$_\odot$, both mass functions are remarkable flat with exponents of -0.77 (Pleiades) and -0.73 (Psc-Eri). Since these slopes are larger than -1, lower-mass stars contribute less than higher mass stars to the total mass of the stream, respectively of the cluster. If the Initial Mass Function (IMF) of Psc-Eri and of the Pleiades followed a Kroupa IMF \citep{2001MNRAS.322..231K} with slopes of  -2.35 for $ m \geq$ 0.5~M$_\odot$ and -1.3 for $ m \leq$ 0.5~M$_\odot$, then both aggregations must have lost the majority of their stars with masses lower than 1~M$_\odot$.

\begin{figure}[hbtp]
   \centering
   \includegraphics[width=0.49\textwidth]{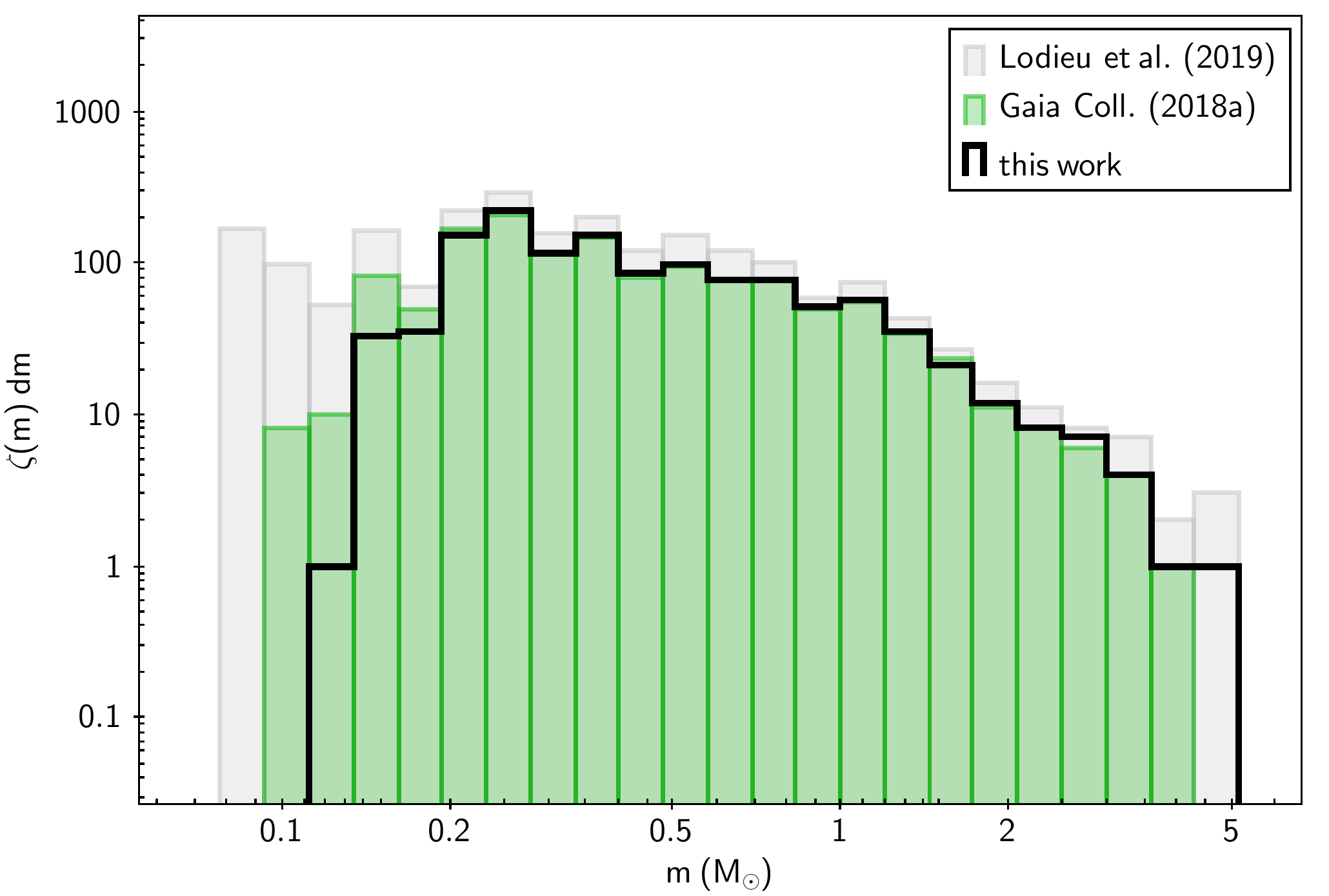}
      \caption{PDMFs of the Pleiades based on recent membership determinations on the basis of Gaia DR2. The PDMF of \citet{2018A&A...616A..10G} is shown  as semi-filled green bars, that from \citet{2019A&A...628A..66L} as semi-filled grey bars, the PDMF from this paper is represented by the black step function.}
         \label{MFcp}
   \end{figure}

While so far there is no other determination of the PDMF of Psc-Eri in this large mass interval to compare with, there has been work on the Pleiades membership determination based on the release of Gaia-DR2 in spring 2018. This comparison is of interest to confirm the almost flat shape of the PDMF in the range 0.3 and 1.0~M$_\odot$. We compared our Pleiades PMDF from Fig.\ref{lf} (right panel) with those derived from the Pleiades membership lists in \citet{2018A&A...616A..10G} (1326 stars) and \citet{2019A&A...628A..66L} (2281 stars). We determined system masses for the Pleiades members of these two samples via the same procedure as described above. This comparison is shown in Fig.\ref{MFcp}. The Pleiades PDMF from our paper is shown as the black step function and the PDMF from \citet{2018A&A...616A..10G} as semi-filled green bars. In the range of interest between 0.2 and 2.5~M$_\odot$, there is complete coincidence between the \citet{2018A&A...616A..10G} PMDF and ours. We note a lack of some 80 stars less massive than 0.2~M$_\odot$. This comes mainly from the strong data quality criteria (see Sec.\ref{detect}), which were not available at the time when \citet{2018A&A...616A..10G}
made their analysis. The PDMF of \citet{2019A&A...628A..66L} is shown in Fig.\ref{MFcp} as dashed semi-filled grey bars. Again, there is good coincidence between \citet{2019A&A...628A..66L}'s mass function and those from \citet{2018A&A...616A..10G} and this paper in the mass range between 0.2 and 2.5~M$_\odot$. However, the \citet{2019A&A...628A..66L} mass function shows an  excess of about 600 stars, mostly less massive than 0.2~M$_\odot$ that are not contained in the other two samples. It turned out that 481 stars of these were excluded from our 2.4\,mas-Catalogue, which means that they have measurements of poorer quality. 

As a summary of this section we point out that the Psc-Eri stream and the Pleiades agree in their total number of stars and their total mass as well as in  their mass- and luminosity functions (Fig.\ref{lf}). They are also of the same age (Fig.\ref{Figure3}). However in their physical appearance they are completely different (see Fig.\ref{Figure2}).  This means that Psc-Eri and the Pleiades must have undergone a completely different formation and/or evolution scenario in the last 130 Myr. This has also been presumed by \citet{2019AJ....158...77C}, when they found that the age Psc-Eri stream is rather 130 Myr than 1 Gyr old as gauged by \citet{2019A&A...622L..13M}.
\subsection{Structure and Kinematics}\label{Kinema}
The drastic difference in spatial structure is highly evident when comparing the PSC-Eri stream with the Pleiades. Whereas Pleiades members
are strongly concentrated within a radius of about 10~pc around the cluster centre and are gravitationally bound, the PSC-Eri stars are spread over a huge area of a nearly cylindrical
shape with length and thickness of about 700 pc and 100 pc, respectively. Moreover, the latter show a patchy distribution with more or less dense 
regions \citep[see also][]{2019A&A...622L..13M}. In Fig.~\ref{Figure2} we indicate these dense regions of the stream by black colour. They include stream members with at least 10 neighbours in 5-D phase space 
(10NB sample, cf. Sec.~\ref{ITT}). We also revealed three rather dense clumps (30NB sample, pink dots in Fig.~\ref{Figure2}) containing stars with more than 30 neighbours.  
But even in these clumps the star density is lower than in the Pleiades by a factor of at least 25. Although the 2D-velocity dispersion of stars in
the clumps is rather small (from 0.23 $\rm{km\,s^{-1}}$ to 0.36 $\rm{km\,s^{-1}}$), the mass density within the clumps is not sufficient for stars to be gravitationally bound. Curiously, at present epoch, the Pleiades (red dots in Fig.~\ref{Figure2}) are located in the vicinity of the Psc-Eri stream. This, however, does not mean that both have been formed in neighbouring molecular clouds, since we measure rather different space velocities for them. Especially, the $V_c$-components of both samples differ by some 25 $\rm{km\,s^{-1}}$.

\begin{figure}[htb!]
  \centering
  \plotone[width=0.49\textwidth]{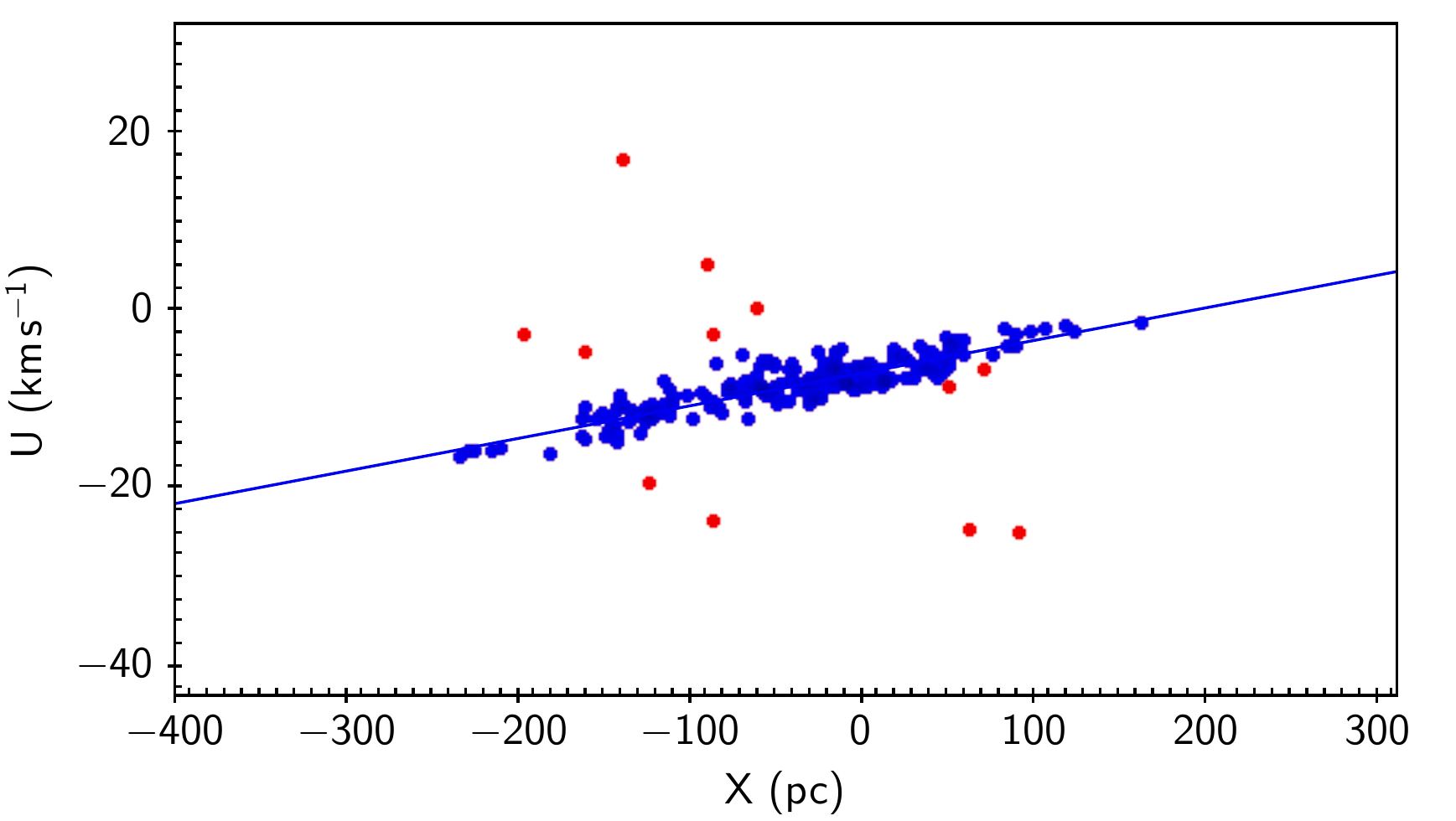}
  \plotone[width=0.49\textwidth]{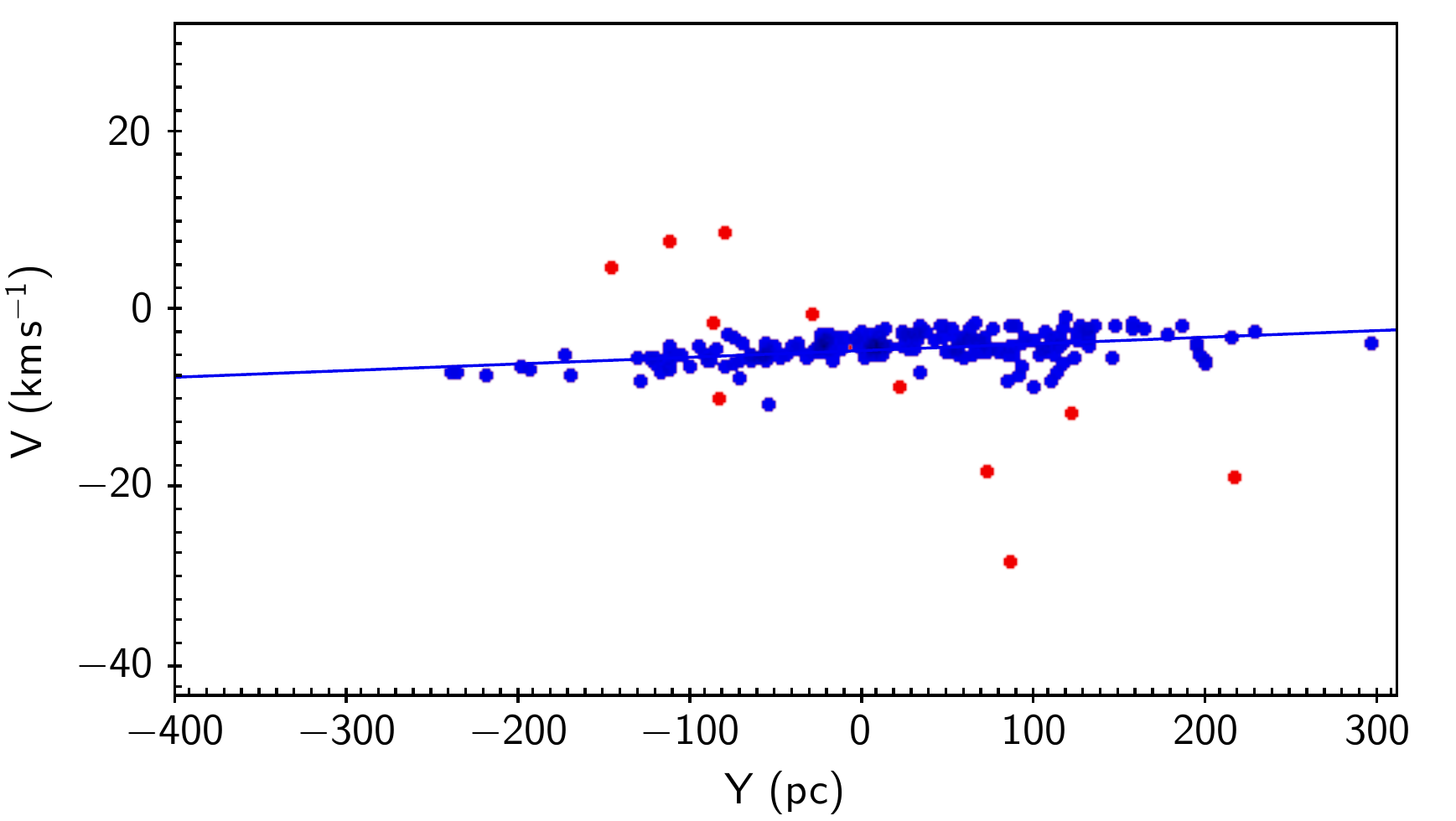}
  \caption{Galactic Cartesian velocities $U,V$ versus coordinates $X,Y$ (from top to bottom) for 205 stars in the psc-Eri stream with radial velocities $V_{rad}$ having $\sigma (V_{rad})$ better than $2.5\,\rm km~s^{-1}$. Blue dots are stars included in the regression analysis, 12 outliers are marked in red. The regression lines are shown in blue.}
  \label{regression}
\end{figure}

In Sec.\ref{CVspace} we already noted correlations between spatial coordinates and velocity components in the Psc-Eri stream. In Fig.~\ref{regression} we show the velocities $U,V$ versus the respective coordinates $X,Y$ based on 205 stars having measurement errors in radial velocity $\sigma (V_{rad})$~<~2.5~km\,s$^{-1}$. Among them, twelve stars show discordant radial velocities (outliers).  
A linear fit between coordinates and velocities delivers slopes of $0.037 $ in $X$ (correlation +0.85), $0.0075$ in $Y$ (correlation +0.4). The formal errors of the slopes are $10^{-6}$. Both slopes are significantly positive in the $X$
 and $Y$-directions which indicates expansion. The expansion times are 27 Myr and 133 Myr in $X$- and $Y$-direction, respectively.
 
The expansion time in $Y$-direction (i.e. direction of Galactic rotation) is remarkably consistent with the age of the Psc-Eri stream found in Sec.\ref{CAMDs}. It is a shear motion similar to that of the tidal tails in the case of the Hyades \citep{2019A&A...621L...2R} or Praesepe \citep{2019A&A...627A...4R}, which are caused by the Galactic gravitational field.

The vertical velocity $W$ of stars perpendicular to the galactic mid-plane is detached from the motion in the plane and is known to be well described by a harmonic oscillator with restoring force. The respective equations are:
\begin{eqnarray}
Z_M(t)& = &Z_M(0) \cos(\nu t) + W_M(0) \nu^{-1} \sin(\nu t),\nonumber \\
W_M(t)    & = &W_M(0)\cos(\nu t) - Z_M(0)\nu \sin(\nu t)\label{oszi}
\end{eqnarray}
where $Z_M$ is the vertical distance from the Galactic mid-plane, $W_M$ is the vertical velocity with respect to the Local Standard of Rest, and $\nu$ is the oscillation frequency. For $\nu$ we adopted $\nu$ = 74 $\rm{km\,s^{-1}\,kpc^{-1}}$ from \citet{2006MNRAS.373..993F} which was based on the local mass density of 0.1 M$_\odot$\,pc$^{-3}$  from \citet{2000MNRAS.313..209H}. The corresponding oscillation period is 84.9 Myr.

For t = 0 we chose the moment of the crossing of the Galactic mid-plane, i.e. $Z_M$(0) = 0. By combining both equations from Eq.~\ref{oszi} we find
\begin{equation}
{\nu}^2{Z_M^2(t)} + {W_M^2(t)} = {W_M^2(0)}.
\end{equation}
The present day vertical distance from the Galactic mid-plane is $Z_M(t) = Z_c + Z_\odot$, where  $Z_\odot = 22$ pc is the distance of the Sun from the mid-plane \citep{2009A&A...495..807K}, hence $Z_M(t)$ = -78.1 pc. The vertical velocity is $W_M(t) = W_c + W_\odot$ with $W_\odot$ = +7.23~km~s$^{-1}$ \citep{2009A&A...495..807K}, hence $W(t)$ = - 11.1~km~s$^{-1}$.
Based on these values we get
${W_M(0)}$ = -12.5~km~s$^{-1}$ and as the maximum height above/below the Galactic mid-plane $\lvert Z_{M,max}\rvert = \pm169$ pc. The last passage through the Galactic mid-plane occurred 6.5 Myr ago, earlier passages took place  49, 91, 134 and 176 Myr (and so on) before present.

It is generally accepted that star formation takes place in regions close to the Galactic mid-plane. As we know from Sec.\ref{CAMDs} the Psc-Eri stream has a formation age window between $\log t = 8.0$ (100 Myr) and 8.2 (158 Myr) before present. If we assume that star formation took place at mid-plane passage, an age of 134 Myr fits best to the age window from the isochrones. Allowing star formation to occur within $\pm 80$~pc around the mid-plane, the age of the Psc-Eri stream  is 134$ \pm 6.5$~Myr. 

For the Pleiades (which are presently also below the Galactic mid-plane) we found the latest mid-plane crossing time to be {-4.5}~Myr. Hence, the Pleiades passed the mid-plane only 2 Myr later than Psc-Eri. Given the oscillation period of 84.9~Myr, both aggregations swing almost in phase.
\section{Origin and Evolution}\label{evo}
From the findings in the previous section we conclude that the Psc-Eri stream is a coherent young stellar aggregation of about 1400 stars moving together and  with a total mass of some 770~M$_\odot$. From this, we expect a common origin of the stars in the same molecular cloud. With respect to total mass and age, Psc-Eri is similar to the open cluster Pleiades, yet they are completely different in their spatial structure. While the Pleiades are a self-gravitationally bound, compact open cluster, the Psc-Eri stream is a huge, gravitationally unbound, bar-like structure. The question arises: why did these two stellar ensembles have such a different fate?

In the past decade, quite some numerical N-body simulations have been performed with the aim of emulating the evolution of open clusters. Details can be found, e.g. in \citet{2018ApJ...863..171S}, where an overview of recent developments in this field is given. Below we only pick up a few aspects of these simulations which can help to answer the aforementioned question.

Star clusters are believed to form during the collapse of giant molecular clouds \citep[see e.g.][]{2019ARA&A..57..227K}. When the first very massive stars (m > 8~M$_\odot$) have formed and evolved, their stellar winds expel the molecular gas from the surroundings of the other newly formed stars, and star formation comes to a hold.

\citet{2017A&A...605A.119S} showed that the star formation efficiency (SFE) in the parental molecular cloud plays a decisive role for the future evolution of the newly formed group of stars. They found that an SFE of at least 15 percent is needed for a newly-born group to survive for more than 100 Myr as a gravitionally bound cluster. Such a value for the SFE has been observed in dense clumps of molecular clouds \citep{2003ARA&A..41...57L,2009ApJ...705..468H}.
A differing SFE may be a first explanation for the different evolution of Psc-Eri and the Pleiades. In the primordial cloud of Psc-Eri, the SFE may have been smaller than 0.15, in that of the Pleiades larger than 0.15, which means the parent cloud of Psc-Eri had a higher mass than that of the Pleiades.


But different values of the SFE may not be the only reason for different development. In a number of numerical N-body simulations \citet{2018ApJ...863..171S} investigated how the stellar mass distribution at the end of molecular gas expulsion influences the later fate of a star cluster. They showed the results for two different stochastic realisations of the same initial conditions. In one case high-mass stars were randomly more concentrated to the centre than in the other. In the first case the life-time as a cluster was longer than in the latter case. In other words, the degree of primordial mass segregation is an important factor for the future development. These simulations \citep{2018ApJ...863..171S} were carried out for high-mass clusters ($ > 3000$~M$_\odot$). The lesson from these numerical experiments is: even with starting conditions describing a high-mass cluster the initial distribution of high-mass stars in the cluster plays an important role for its further evolution. The more this should hold for lower-mass cases as Psc-Eri and the Pleiades. So, it could well be that the different evolutionary history of the Pleiades and Psc-Eri was determined by the primordial distribution of high-mass stars, i.e. Psc-Eri had a lower value of mass segregation than had the Pleiades. It would be worthwhile having N-body experiments for lower mass cases to actually test our conclusions.
 

%
\section{Summary}\label{summ}
We used a modified 5-D convergent point approach to search for hitherto unrevealed members of the Psc-Eri stellar stream detected by \citet{2019A&A...622L..13M}. Based on  Gaia DR2 we found 1387 co-moving members in a G magnitude range from 5.1  mag to G = 19.3 mag.  The stars of the stream are spread over a huge area of nearly cylindrical shape with length and thickness of about 700 pc and 100 pc, respectively. The distance of the stars in the stream varies between  80 pc and 380 pc from the Sun. Depending on the distance, the faint completeness limit is found to range from $M_G$ = 13.5mag to 10.1 mag. The Psc-Eri stream has a patchy structure with clumps of higher space density of up to 40 stars within a sphere of 10~pc radius. Even in these regions, stars are gravitationally unbound, as is the stream as a whole. The stream is presently 78 pc below the Galactic mid-plane, and will reach a maximum distance from the mid-plane of about 170~pc.

The loci of the Psc-Eri members in the $G$ vs. $(G - G_RP)$ CAMD are consistent with PARSEC (version 1.2S) solar metallicity isochrones \citep{2014MNRAS.444.2525C} for ages between 100 ($\log \rm{t}$ = 8.0) and 158 Myr ($\log \rm{t}$ = 8.2). The absence of absolutely bright stars does not allow to narrow this time window. An independent age indicator is provided by the oscillation period of the stream perpendicular to the Galactic mid-plane. We found a value of 134 $\pm 6.5$ Myr as the age most compatible with the isochronic age.   

The stream is expanding both in the direction of Galactic rotation $Y$, as well as in the direction towards the Galactic centre $X$, the corresponding expansion time-scales are 133 Myr and 27 Myr. Again, the expansion time in the $Y$ direction is in agreement with the age of the cluster. Another independent age estimate was already obtained by \citet{2019AJ....158...77C} who derived an age of about 120 Myr from stellar rotation period measurements with the TESS satellite, using a comparison of the locations of Psc-Eri stars and Pleiades stars in the  (Rotation Period-Effective Temperature)-diagram. 

Indeed, we found that the Psc-Eri stream and the Pleiades cluster are very similar in many astrophysical characteristics, not only with regard to their age.  Both stellar aggregations follow the same luminosity function in $M_G$, hence the same present-day mass function. The PDMF  $\zeta(m)$  between 0.3 and 2.5~M$_\odot$ is well represented by a broken power law $\zeta(m) \propto \rm{m}^{-\alpha}$ with $\alpha = 2.15$ for masses between 1.0 and 2.5~M$_\odot$ and $\alpha = 0.73$ for masses between 0.3 and 1.0~M$_\odot$ for Psc-Eri. The corresponding exponents for the Pleiades are 2.41 and 0.77, respectively. While both mass functions for masses higher than 1~M$_\odot$ almost follow a Salpeter law with $\alpha = 2.35$, they are very shallow and even top-heavy below 1~M$_\odot$. If their IMFs had a shape as in \citet{1955ApJ...121..161S} or \citet{2001MNRAS.322..231K}, the shallow slope indicates high mass-loss in the regime of low-mass stars. The total mass of the 1387 members of the Psc-Eri stream adds up to 771~M$_\odot$. That is very similar to the total mass of the Pleiades within two Jacobi radii of 690~M$_\odot$ in 1245 stars. At its age of 134 Myr, the stream hosts 3 white dwarfs.

Although the Pleiades and the Psc-Eri stream resemble each other in many astrophysical properties, their shapes are completely different. Curiously enough, at present the minimum distance between the Pleiades and Psc-Eri is only 80~pc. As their space velocities differ by about 25~km~s$^{-1}$, they were much farther away from each other in the past and formed in two different molecular clouds. The results of numerical N-body simulations recently carried out by \citet{2017A&A...605A.119S} and \citet{2018ApJ...863..171S} suggest two different possibilities to explain the different fate of Psc-Eri and the Pleiades: the star formation efficiency in the Psc-Eri parental cloud was lower than in that of the Pleiades or - given the same SFE - mass-segregation after gas expulsion in both groups was higher in the Pleiades than was in Psc-Eri.  
\begin{acknowledgements}
It is a pleasure to thank Andreas Just and Peter Berczik of ZAH, University of Heidelberg, for helpful discussions, especially on the evolution of Galactic star clusters.
This research has made use of the SIMBAD database and the VizieR catalogue access tool, operated at CDS, Strasbourg, France.
This work has made use of data from the European Space Agency (ESA)
mission \gaia\ (\url{https://www.cosmos.esa.int/gaia}), processed by
the \gaia\ Data Processing and Analysis Consortium (DPAC,
\url{https://www.cosmos.esa.int/web/gaia/dpac/consortium}). Funding
for the DPAC has been provided by national institutions, in particular
the institutions participating in the \gaia\ Multilateral Agreement.
\end{acknowledgements}
\bibliography{mybib}
%
%

\clearpage

\begin{appendix}\label{sec:appen}
\section{Tables}
\tiny
\begin{table*}[h!]
\begin{tabular}{r r r r r r r r r r r r}
\hline\hline
  \multicolumn{1}{c}{Gaia DR2 source\_id} &
  \multicolumn{1}{c}{RA} &
  \multicolumn{1}{c}{Dec} &
  \multicolumn{1}{c}{G} &
  \multicolumn{1}{c}{$X$} &
  \multicolumn{1}{c}{$Y$} &
  \multicolumn{1}{c}{$Z$} &
  \multicolumn{1}{c}{$\Delta V_{\parallel}$} &
  \multicolumn{1}{c}{$\Delta V_{\bot}$} &
  \multicolumn{1}{c}{$\log$} &
  \multicolumn{1}{c}{mass} &
  \multicolumn{1}{c}{RUWE} \\
  \multicolumn{1}{c}{} &
  \multicolumn{1}{c}{deg} &
  \multicolumn{1}{c}{deg} &
  \multicolumn{1}{c}{mag} &
  \multicolumn{1}{c}{pc} &
  \multicolumn{1}{c}{pc} &
  \multicolumn{1}{c}{pc} &
  \multicolumn{1}{c}{km~s$^{-1}$} &
  \multicolumn{1}{c}{km~s$^{-1}$} &
  \multicolumn{1}{c}{$P_{conta}$} &
  \multicolumn{1}{c}{M$_{\odot}$} &
  \multicolumn{1}{c}{} \\
\hline
  2341740796846600576 & 0.19432 & -20.48931 & 15.285 & 11.34 & 19.70 & -95.83 & -0.60 & 0.01 & -44.30 & 0.35 & 1.16\\
  2448628204833165056 & 0.34745 & -2.48618  & 16.155 & -5.06 & 59.29 & -114.79 & 0.68 & 0.91 & -16.77 & 0.35 & 0.99\\
  2745159000421889792 & 0.42684 & 5.84435   & 10.975 & -13.34 & 68.22 & -98.91 & 1.60 & -0.98 & -22.25 & 0.90 & 1.12\\
  2745207958752381824 & 0.53167 & 6.15672   & 18.797 & -14.42 & 71.46 & -102.73 & 0.81 & 0.88 & -36.03 & 0.14 & 0.96\\
  2415762324810035200 & 0.75821 & -16.20197 & 16.279 & 6.54 & 23.92 & -88.45 & -1.18 & 0.41 & -38.03 & 0.25 & 1.01\\
  2340695294431217536 & 0.88062 & -21.39156 & 18.091 & 11.26 & 18.08 & -97.48 & -0.46 & -0.07 & -40.10 & 0.14 & 0.91\\
  2444417315816670336 & 0.92757 & -5.45925  & 15.816 & -2.93 & 53.48 & -117.70 & 0.27 & 2.16 & -3.00 & 0.40 & 1.09\\
  2738830452009692544 & 1.04401 & 1.75111   & 15.230 & -11.41 & 67.85 & -114.31 & 0.43 & 0.56 & -36.03 & 0.45 & 1.18\\
  2738830452009692672 & 1.04572 & 1.75006   & 16.339 & -11.53 & 68.56 & -115.51 & 0.23 & 0.33 & -36.03 & 0.30 & 1.05\\
  2741665405303366912 & 1.18058 & 4.28959   & 10.531 & -13.11 & 64.99 & -100.62 & 0.31 & -0.84 & -30.06 & 1.0 & 0.79\\
\hline\end{tabular}
\caption{The first ten stars of our 1387 Psc-Eri stream members. The full table  is available online via CDS.}
    \label{tab:streamsources}
\end{table*}
\begin{table*}[h!]
\begin{tabular}{r r r r r r r r r r r r}
\hline\hline
  \multicolumn{1}{c}{Gaia DR2 source\_id} &
  \multicolumn{1}{c}{RA} &
  \multicolumn{1}{c}{Dec} &
  \multicolumn{1}{c}{G} &
  \multicolumn{1}{c}{$X$} &
  \multicolumn{1}{c}{$Y$} &
  \multicolumn{1}{c}{$Z$} &
  \multicolumn{1}{c}{$\Delta V_{\parallel}$} &
  \multicolumn{1}{c}{$\Delta V_{\bot}$} &
  \multicolumn{1}{c}{$\log$} &
  \multicolumn{1}{c}{mass} &
  \multicolumn{1}{c}{RUWE} \\
  \multicolumn{1}{c}{} &
  \multicolumn{1}{c}{deg} &
  \multicolumn{1}{c}{deg} &
  \multicolumn{1}{c}{mag} &
  \multicolumn{1}{c}{pc} &
  \multicolumn{1}{c}{pc} &
  \multicolumn{1}{c}{pc} &
  \multicolumn{1}{c}{km~s$^{-1}$} &
  \multicolumn{1}{c}{km~s$^{-1}$} &
  \multicolumn{1}{c}{$P_{conta}$} &
  \multicolumn{1}{c}{M$_{\odot}$} &
  \multicolumn{1}{c}{} \\
\hline
  111728390182594304 & 46.93316 & 24.84404 & 16.551 & -105.53 & 42.04 & -61.71 &  0.64 & -0.96 & -25.46 & 0.30 & 0.93\\
  111721552594418304 & 47.40077 & 24.90663 & 12.123 & -103.83 & 40.62 & -59.83 & -0.43 & -0.75 & -45.00 & 0.75 & 0.98\\
  112256980396927360 & 47.96920 & 25.84448 & 9.3210 & -116.95 & 46.06 & -64.48 &  0.07 &  2.14 &  -0.31 & 1.30 & 0.88\\
  112256976100897408 & 47.97014 & 25.84525 & 12.692 & -117.18 & 46.15 & -64.60 &  0.08 &  0.96 & -45.00 & 0.74 & 1.23\\
  122441859683763712 & 48.54406 & 29.36330 & 15.295 & -109.03 & 46.87 & -52.79 & -0.50 &  0.04 & -45.00 & 0.44 & 1.17\\
  118289931977488512 & 48.86221 & 26.29361 & 10.464 & -118.01 & 45.43 & -62.57 & -0.32 &  2.33 &  -0.31 & 1.05 & 1.10\\
  118715683496508032 & 48.90396 & 27.77819 & 15.507 & -106.89 & 43.14 & -53.99 & -1.48 &  0.65 & -45.00 & 0.42 & 1.18\\
  62413988007668352  & 49.45744 & 22.83200 & 7.5640 & -112.71 & 37.15 & -65.21 &  0.20 & -0.24 & -45.00 & 1.92 & 0.92\\
  117448530705462912 & 49.67108 & 25.78522 & 15.743 & -118.12 & 43.15 & -62.18 & -0.30 & -0.16 & -45.00 & 0.42 & 1.18\\
  62104028807382784  & 50.31235 & 21.85965 & 16.081 & -104.49 & 31.64 & -60.70 &  0.74 & -0.70 & -45.00 & 0.35 & 1.07\\
\hline\end{tabular}
\caption{The first ten stars of our 1245 Pleiades members. The full table  is available online via CDS.}
    \label{Pleiades}
\end{table*}

\end{appendix}

\end{document}